\documentclass[prl,superscriptaddress,showpacs,preprintnumbers,amsmath,amssymb]{revtex4}
\usepackage[ansinew,latin1]{inputenc}
\usepackage{psfrag}
\usepackage{array}
\usepackage{amssymb}
\usepackage{amsmath}
\usepackage{graphicx}
\usepackage{xcolor}

\begin{document}

\title{Is quantum nonlocality definitively confirmed by the experiments ?}

\author{M. Iannuzzi}
\email{mario.iannuzzi@roma2.infn.it}
\thanks{Corresponding author}
\affiliation{Dipartimento di Fisica, Universit\`a di Roma ``Tor Vergata", I-00133 Roma, Italy}

\date{\today}


\begin{abstract}
In a  recent paper by B.Hensen {\it et al.}, {\it Nature} {\bf 526}, 682 (2015)  and in two companion papers 
by M. Giustina {\it et al.}, {\it Phys. Rev. Lett.} {\bf 115}, 250401 (2015) and by L. Shalm {\it et al.} 
{\it Phys. Rev. Lett.} {\bf 115}, 250402 (2015), the authors describe beautiful and complex experiments 
aimed at testing the theorem of J. Bell, {\it Physics} (NY), {\bf 1}, 195 (1964)  with measurements free 
of the detection-loophole and the locality loophole that had never been closed unquestionably in a 
single experiment. According to their authors, each experiment, closing both  loopholes, gives further and
stronger evidence in favor of nonlocality. Nevertheless, I will use this occasion to argue that an ispection,
merely of experimental/descriptive nature and independent of any theoretical assumption, of the conceptul
scheme of measurement of the Bell-type experiments performed to date shows that likely all of the experimental
requirements for testing Bell's theorem are not actually satisfied in the laboratory reality. Hence, new
experiments should close also the ``measerument loophole'' here indicated.
\end{abstract}

\pacs{03.65.Ud 03.65.Ta}

\maketitle

\section{Introduction}
The experiments described in Ref. \cite{ref1,ref2,ref3} are tests of Bell's theorem performed with measurements free of
loopholes. Their experimental configurations are different : in experiment \cite{ref1} entangled pairs of particles were
created at two independent sources at spacelike separation; in experiments  \cite{ref2} 
and \cite{ref3}  entangled pairs of particles (photons) were emitted from the same source, and  two detectors, each recording 
one particle, were at spacelike separation. Furthermore, in experiment \cite{ref1} entaglement was generated differently 
from experiments \cite{ref2} and \cite{ref3}. Anyway my comment will concern the general features of all of the Bell-type
experiments, independently of the  configuration of a specific apparatus. First, let me recall their conceptual scheme of
measurement.
Consider a two-component physical system formed of two entangled particles, and two instruments (consisting of polarizer 
and detector), each  associated with one component. The detectors merely record counts. Each instrument has a range of 
settings, denoted as {\bf a} for the first instrument and {\bf b} for the second instrument. The physical quantities that will 
be measured  are the joint probabilities of simultaneous detection of the two particles. The results of a measurement depend 
also on uncontrolled parameters $\lambda$. A Bell-like inequality, like the one derived by CHSH \cite{ref5} or by CH \cite{ref6}, 
is always satisfied by  physical systems formed of pairs of particles whose components are spatially separated and independent of
each other (locality). The basic mathematical form of the CH inequality, upon averaging over the 
$\lambda$-distribution, is  : 

\begin{equation}
\frac{p_{12} ({\bf a}, {\bf b}) - p_{12} ({\bf a}, {\bf b'}) + p_{12} ({\bf a'}, {\bf b}) + p_{12} ({\bf a'}, {\bf b'})}
{p_1({\bf a'})  +  p_2({\bf b})}   \le  1                               
\label{eq1}                  
\end{equation}

\noindent where $p_{12} ({\bf a}, {\bf b})$ is the probability of detecting both components of the pair, and $p_1({\bf a})$  and 
$p_2({\bf b})$ are single-count probabilities. Unlike inequality (\ref{eq1}), if  each detector, instead of merely recording
counts, were to detect two possible results, like +1 and -1, the inequality is:

\begin{equation}
E ({\bf a}, {\bf b}) - E ({\bf a}, {\bf b'}) + E ({\bf a'}, {\bf b}) + E ({\bf a'}, {\bf b'})  \le  2                         
\label{eq2}                  
\end{equation}

\noindent where $E ({\bf a}, {\bf b}) = p_{++}({\bf a}, {\bf b}) + p_{--} ({\bf a}, {\bf b}) - p_{+-}({\bf a}, {\bf b}) -
p_{-+} ({\bf a}, {\bf b})$ is the correlation function defined by Bell.  $p_{++} ({\bf a}, {\bf b})$  is the joint probability
of both instruments {\bf a} and {\bf b} detecting +1, $p_{+-} ({\bf a}, {\bf b})$ is the joint probability of the first instrument
recording +1 and the second instrument recording -1, and so on. However, in Appendix B of \cite{ref6} it is shown that inequality
(\ref{eq2}) is a corollary of inequality (\ref{eq1}), so that any conceptual comment on (\ref{eq1}) {\it is also applicable} to
(\ref{eq2}). Famous experiments testing these inequalities were performed on pairs of entangled optical photons [7$\div$10], and
recently the first experiment on pairs of spin-1/2 particles in the singlet state  has been reported \cite{ref11}, all
contradicting the local theories.

\section{Role of conditional probabilities in a Bell-type experiment}

My comment is quite simple. A Bell-type experiment measures, with different settings of the apparatus, the joint probabilities
of detecting one particle of the entangled pair {\it only if}  the other particle has also been detected (simultaneously, or
within the time-window and the time-delay specific of each experiment), this condition being the prerequisite of a coincidence
measurement. Consequently, {\it and independently of any theoretical/interpretative approach}, the joint probability of
entangled-particles' transmission through the polarizers will have the form

\begin{equation}
p_{12}({\bf a}, {\bf b})=p({\bf a},1)\cdot p({\bf b},2|{\bf a},1)
\label{eq3}                  
\end{equation}

\noindent where $p({\bf b},2|{\bf a},1)$ is the {\it conditional probability} of detecting particle-2 at detector-2 {\it if}
particle-1 has been detected at detector-1. Clearly in this equation the observable $p({\bf b},2|{\bf a},1)$ {\it is not local :
its value depends on the final preparation-state of the entangled pair of spatially-separated photons after the polarizers,
which in turn depends on the setting of polarizers-2 (at the crossing time of particle-2) relative to the setting of the
spatially-separated polarizer-1 (at the crossing time of particle-1).} Then, as a result of the adopted experimental scheme
in which the unique final outcome is obtained from the measurement of conditional probabilities, the measurement of either
particle of an entangled pair is necessarely performed ({\it via} $p({\bf b},2|{\bf a},1)$) in association with the detection
of the other particle, and so will not be independent of what other measuremts must be done simultaneously on the latter.
Note that the present comment is merely of experimental nature. Its conceptual implications on the experimental tests of Bell's
inequalities will be commented below.

Inspection of inequality (\ref{eq1}) shows clearly that the experimental conditions for {\it possible} occurrence of 
violations are those determining particular behaviors of $p_{12}({\bf a}, {\bf b})$ which also include, at certain settings of the 
apparatus, values {\it larger than the probability of coincidence of two independent particles}. Since, for certain states of the 
physical system and particular experimental arrangements, such circumstancies can definitely be verified, from (\ref{eq3}) 
violations of  inequality (\ref{eq1}) may occur.
Note that the present examination of a possible origin of violations of the CH inequality implies no need 
of any information (about the analyzers' settings) traveling from one observer to the other, and, more in general, no need for 
invoking nonlocality of the observations : the occurrence of violation is determined solely by the final preparation state of the 
entangled physical system (that is, the final state on which the coincidence measurement is done ) and by {\it the actual 
configuration of the apparatus which can be changed by independent manipulation of each polarizer}. According to this 
reflection, in Bell-type experiments the distance between the observers is conceptually irrelevant, and no locality loophole is 
to be closed. Here and in the following, {\it I will knowingly use the concept of "nonlocality", and not the concept of "violation 
of the local realist theories".} With the wording of L. Ballentine \cite{ref13}, the term "local {\it realism}", very often
used in relation to the conclusions of the Bell theorem, is "seriously misleading, ... for, without some degree of {\it realism}
the distinction between {\it local} and {\it nonlocal} has no meaning, ... and a local {\it non-realistic} theory is an empty set".
So I will use the expression "local theory" instead of "local realist theory" because "some degree of realism" is inherent in the
concept of locality. The exclusion of the superfluous adjective "realist" is not a drop of realism of the local theories; it is
just the opposite. Clearly, this choice is absolutely non influential in my argument of experimental/descriptive nature.

A clear exemplification of the approach above is given by a Bell-type optical polarization experiment using {\it linear}
polarizers. In such an experiment, the entangled particles are two optical photons with parallel linear polarization, each of
which may pass through a linear polarizing filter before reaching a detector merely recording counts. The first filter,
associated with the first photon, is oriented in the {\bf a} direction, and the other filter, associated with the second photon,
is oriented in the {\bf b} direction in the $xy$ plane. According to the discussion above, the coincidence probability of
simultaneous detection of both photons transmitted through the linear filters is:

\begin{equation}
p({\bf a},1)\cdot p({\bf b},2|{\bf a},1)=\frac{1}{2}\cos^2(\vartheta_{ab})
\label{eq4}                  
\end{equation}
                                    
Here, $\vartheta_{ab}$  is the angle between the polarizers' axes, and  $p ({\bf a},1) = 1/2$ is the probability of the first 
photon passing through the first linear polarizer oriented in the {\bf a} direction.  $p({\bf b},2|{\bf a},1) =
\cos^2(\vartheta_{ab})$  is the {\it conditional} probability of the second photon passing through the second linear polarizer
oriented in the {\bf b} direction, {\it if} the first photon passes through the first filter oriented in the {\bf a} direction;
in fact, according to quantum theory, such probability is  given by the squared projection along {\bf a} of photon-2 exiting
filter-2 with polarization direction  {\bf b}. Due to the invariance of the states under rotation about the z axis, it is
independent of the absolute direction of {\bf a} and {\bf b}, and depends only on the relative angle $\vartheta_{ab}$.
Alternatively, this result is generally calculated from the state vector  $|\Psi\rangle$ of the two entangled particles :    

\begin{equation}
p({\bf a},1)\cdot p({\bf b},2|{\bf a},1)\equiv p_{++}({\bf a}, {\bf b})=\langle \Psi|\delta_1\delta_2 |\Psi\rangle=\frac{1}
{2}\cos^2(\vartheta_{ab})
\label{eq5}                  
\end{equation}
              
\noindent where  $\delta_i$  ( i  = 1, 2) is  the operator projecting the polarization vector of particle-i on the axis of the
linear polarizer-i. 
Now, if we substitute into inequality (\ref{eq1}) the ideal  predictions for the state  $|\Psi\rangle$,   
$p({\bf a},1)\equiv p_1({\bf a})=1/2$, $p({\bf b},2)\equiv p_2({\bf b})=1/2$,  $p_{++}({\bf a}, {\bf b})=p_{12}(\vartheta_{ab})=
(1/2) \cos^2(\vartheta_{ab})$ and the polarization directions generally chosen to test the C-H inequality,  the predicted angular range of  violations is  $0 <  \vartheta_{ab} < \pi /4$, with the maximum 
violation occurring  at  $\vartheta_{ab}= \pi /8$, corresponding to values of $p({\bf b},2|{\bf a},1) = \cos^2(\vartheta_{ab})$  
higher than the single-count probability $p_2({\bf b}) = 1/2$ \cite{ref12}. 
The experiments have fully confirmed the prediction. This shows that, for certain settings of the two polarizers, the final states
of the entangled physical system correspond to values of the conditional probability $p({\bf b},2|{\bf a},1)$  higher than the
value  $p({\bf b},2)$
representing the absence of particle-1. In such circumstances, violations of the Bell theorem may occur, but their interpretation 
does not imply any concept of nonlocality of the observations. This conclusion is compatible with both the result  derived from
theoretical analyses with observable algebras, which conclude that "the nonlocality of entangled states and the locality of
observable algebras are two perfectly compatible aspects of quantum physics" \cite{ref15}, and particularly an old suggestion 
of J. Bell \cite{ref14} who pointed out that in the derivation of the no hidden-variables theorems it is " tacitly assumed that
the measurement of an observable must yield the same value independently of what other measurements must be done simultaneously",
whereas he suggested that " the result of an observation may reasonably depend not only on the state of 
the system (including hidden variables) but also on the complete disposition of the apparatus".

\section{Conclusion} 

The present comment is merely an experimental/descriptive inspection of the measuring scheme of a Bell-type experiment,
independent of any theoretical assumption, and it can be summarized as follows :
\begin{itemize}
\item[a)] in the laboratory reality, the final result is obtained from the measurement of the conditional probability
  $p({\bf b},2|{\bf a},1)$ of detecting either particle of an entangled pair, which in turn yields the coincidence
  probability $p_{12}({\bf a}, {\bf b}) = p({\bf a},1) \cdot p({\bf b}, 2 | {\bf a},1)$; 
\item[b)] the conditional probability is a nonlocal observable, and its value is not independent of what measurements
  are done simultaneously on the other particle by manipulation of the corresponding  polarizer;
\item[c)]  from (a) and (b) the measurements done in the experimental tests  performed to date do not seem to satisfy the
assumption, of the no hidden-variables theorems indicated by J. Bell, and they may show violations merely due to
the polarizers settings, i.e. to the complete disposition of the apparatus. In particular, this circumstance may occur
for certain behaviors of the conditional probabilities with some of their values satisfying the criterion
$p({\bf b}, 2|{\bf a},1) > p({\bf b}, 2)$.
\end{itemize}

This result, showing that an observation may indeed show violation of Bell theorem due to the experimental arrangement, seems
to confirm  that not all of the requirements of Belltheorem  have been actually satisfied in the real experiments, which
therefore cannot test alone local causality. New experiments should then be done to close also the ``measurement loophole''
here discussed.

The above remarks have concerned Bell-like inequalities restricting correlations or probabilities. The author is well aware that 
Bell's theorem has also been obtained without using probabilities or inequalities, with a derivation showing that the assumption of 
noncontextual values of quantum-mechanical observables leads to a contradiction \cite{ref16, ref17, ref18}. Moreover, it has 
been shown that for certain prepared states the noncontextuality of the measured values is directly derived from locality. 
It seems therefore plausible that the question of the nonlocal nature of the quantum theory should be experimentally  tested and 
closed only in this context.

\section{Acknowledgments}
The author is indebted to D. Moricciani for helpful comments.

\end{document}